\documentclass[aps,superscriptaddress,reprint]{revtex4-2}
\usepackage{amsmath}
\usepackage{amssymb}
\usepackage{mathrsfs}
\usepackage{graphicx}
\usepackage{booktabs}  
\usepackage{multirow}  
\usepackage{natbib}    
\usepackage[colorlinks,citecolor=blue, linkcolor=blue,hyperindex,CJKbookmarks,]{hyperref}
\hypersetup{colorlinks, citecolor=magenta, linkcolor=magenta, urlcolor=magenta,hyperindex,CJKbookmarks,}
\begin{document}
	\title{Purely Quantum Nonreciprocity by Spatially Separated Transmission Scheme}
	
	\author{Zhi-Hao Liu}
	\affiliation{Key Laboratory of Low-Dimensional Quantum Structures and
		Quantum Control of Ministry of Education, Key Laboratory for Matter
		Microstructure and Function of Hunan Province, Department of Physics and
		Synergetic Innovation Center for Quantum Effects and Applications, Hunan
		Normal University, Changsha 410081, China}
	
	\author{Guang-Yu Zhang }
	\affiliation{Key Laboratory of Low-Dimensional Quantum Structures and
		Quantum Control of Ministry of Education, Key Laboratory for Matter
		Microstructure and Function of Hunan Province, Department of Physics and
		Synergetic Innovation Center for Quantum Effects and Applications, Hunan
		Normal University, Changsha 410081, China}
	
	\author{Xun-Wei Xu}
	\email{xwxu@hunnu.edu.cn}
	\affiliation{Key Laboratory of Low-Dimensional Quantum Structures and
		Quantum Control of Ministry of Education, Key Laboratory for Matter
		Microstructure and Function of Hunan Province, Department of Physics and
		Synergetic Innovation Center for Quantum Effects and Applications, Hunan
		Normal University, Changsha 410081, China}
	\affiliation{Institute of Interdisciplinary Studies, Hunan Normal University, Changsha, 410081, China}
	
	\date{\today}
	
	\begin{abstract}
		Nonreciprocal photon blockade is of particular interest due to its potential applications in chiral quantum technologies and topological photonics. In the regular cases, nonreciprocal transmission (classical nonreciprocity) and nonreciprocal photon blockade (quantum nonreciprocity) often appear simultaneously. Nevertheless, how to achieve \emph{purely} quantum nonreciprocity (no classical nonreciprocity) remains largely unexplored. Here, we propose a spatially separated transmission scheme, that the photons transport in different directions take different paths, in an optical system consisting of two spinning cavities coupled indirectly by two common drop-filter waveguides. Based on the spatially separated transmission scheme, we demonstrate a purely quantum nonreciprocity (nonreciprocal photon blockade) by considering the Kerr nonlinear interaction in one of the paths. Interestingly, we find that the nonreciprocal photon blockade is \emph{enhanced nonreciprocally}, i.e., the nonreciprocal photon blockade is enhanced when the photons transport in one direction but suppressed in the reverse direction.
		We identify that the nonreciprocal enhancement of nonreciprocal photon blockade is induced by the destructive or constructive interference between two paths for two photons passing through the whole system.
		The spatially separated transmission scheme proposed in the work provides a novel approach to observe purely quantum nonreciprocal effects.
	\end{abstract}
	\maketitle
	
	\section{Introduction}
	
	
Optical nonreciprocal transmission is an interesting phenomenon that photons are allowed to pass in one direction but blocked in the opposite direction.
It is pivotal for realizing unidirectional optical devices, including optical isolators~\cite{Jalas2013NaPho,Sayrin2015,XuXW2018PRA,Tang2019,TangL2022PRL,NieW2022SCPMA}, circulators~\cite{XuXW2015PRA,Ruesink2018NatCo,XiaK2018PRL,LiE2020PRR,YanW2020OPTICA}, directional amplifiers~\cite{Metelmann2015PRX,LiY2017OExpr,Malz2018,Shen2018,SongLN2019PRA,JiangC2019PRA,ZhangXZ2018PRA,Mercier2019PRAPP,Mercier2020PRL}, unidirectional frequency converter~\cite{XuXW2016PRA,XUXW2017PRA,Peterson2017,Bernier2017NatCo,Barzanjeh2017NatCo,LiG2018PRA,DuL2021PRR}, etc.
By breaking Lorentz reciprocity~\cite{Sounas2017}, optical nonreciprocity has been observed based on the magneto-optic effect~\cite{BiL2011NaPho,Shalaby2013NatCo,Shoji_2014}, optical nonlinearity~\cite{Fan2012Sci,sounas2018broadband,YangPF2019PRL}, synthetic magnetism~\cite{Fang2012NaPho,Tzuang2014NaPho,Fang2017NatPh,LuX2021PRL}, indirect interband transition~\cite{YuZ2009NaPho,Lira2012PRL,Castellanos2014PRL,Tian2021NaPho}, parity-time symmetry breaking~\cite{Ruter2010NatPh,PengB2014NatPh,ChangL2014NaPho,Ramezani2014PRL}, Doppler effect~\cite{WangDW2013PRL,Horsley2013PRL,Zhang2018NaPho,Maayani2018Natur}, chiral interaction~\cite{Scheucher2016Sci,Lodahl2017Natur,TangJS2022PRL}, and so on.
	
As an important development direction, the concept of optical nonreciprocity was extended from the classical regime to quantum regime, and various of quantum nonreciprocal effects were predicted, e.g., nonreciprocal photon blockade~\cite{HuangR2018PRL,Xu2020PRJ,XiaX2021PRA,XieH2022PRA,ShenC2023PRA},  nonreciprocal entanglement~\cite{JiaoYF2020PRL,JiaoYF2022PRAPP,Ren2022OL,ChenJ2023PRB}, and nonreciprocal squeezing~\cite{ChenSS2021ADP,Zhao2024OL,GuoQ2023PRA,wang2024quantum}.
For example, nonreciprocal photon blockade is referred to the phenomenon that photon blockade happens when the cavity is driven in one direction but not the other, which opens a route for chiral quantum manipulation of light.
In a recent work, chiral cavity quantum electrodynamics system is experimentally demonstrated with multiple atoms strongly coupled to a Fabry–Pérot cavity~\cite{YangPF2023LPR}.
	
Different from classical nonreciprocity, nonlinear interaction is an essential prerequisite to nonreciprocally manipulate the quantum properties of photons.
One of the most popular approaches for quantum nonreciprocity is to introduce nonlinear interactions in the classically nonreciprocal systems, such as spinning resonators with additional Kerr nonlinearity~\cite{HuangR2018PRL,JingYW2022NatSR}, second-order nonlinearity~\cite{WangK2019PRA,ShenHZ2020PRA,GouC2023PRA}, optomechanical interaction~\cite{LiBJ2019PRJ,LiuYM2023OE,Shang_2021}, or atom-cavity interaction~\cite{XueW2020OL,JingYW2021PRA,Wang_2021EPL,LiuYM2023PRA,ZhangW2023SCPMA}.
In such cases, nonreciprocal transmission (classical nonreciprocity) and nonreciprocal photon blockade (quantum nonreciprocity) often appear simultaneously. 
Here, we consider another interesting quantum nonreciprocity that nonreciprocal photon blockade is achieved without classical nonreciprocity, referred to as \emph{purely} quantum nonreciprocity.
Except for a few exceptions~\cite{XuXW2020PRAPP,GrafA2022PRL,XiangY2023PRA}, purely quantum nonreciprocity is hard to achieve based on the popular approaches for quantum nonreciprocity, because the paths for the photon transmission are spatially unseparated, i.e., the photons travel in opposite directions but in the same resonator (path).

In this paper, we propose a spatially separated transmission scheme to achieve purely quantum nonreciprocity, in an optical system consisting of two spinning cavities coupled indirectly by two common drop-filter waveguides. 
We show that the photons transport between two ports with the same transmission rate (classical nonreciprocity), but photons pass through different paths (cavities) when they transport in different directions.
Due to the Kerr nonlinear interaction in one of the paths, we demonstrate nonreciprocal photon blockade (purely quantum nonreciprocity) via the spatially separated transmission scheme. 
Surprisingly, we find that the nonreciprocal photon blockade is enhanced nonreciprocally, i.e., the nonreciprocal photon blockade is enhanced when the photons transport in one direction but suppressed in the reverse direction.
We identify that the nonreciprocal enhancement of nonreciprocal photon blockade is induced by the destructive or constructive interference between two paths for two photons passing through the whole system.
Spatially separated transmission scheme provides a novel approach for purely quantum nonreciprocity.

This paper is organized as follows: In Sec.~\ref{sec2},  we introduce a physical model consisting of a spinning Kerr cavity and a spinning linear cavity coupled indirectly by two common drop-filter waveguides. In Sec.~\ref{sec3}, we explore the spatially separated transmission scheme and show the paths of the photons when they transport in different directions.
In Sec.~\ref{sec4}, we demonstrate the nonreciprocal photon blockade and its nonreciprocal enhancement effect based on numerical simulations. 
The mechanism for the nonreciprocal enhancement of nonreciprocal photon blockade is analyzed analytically in Sec.~\ref{sec5}. 
Finally, a conclusion is given in Sec.~\ref{sec6}.
	
\section{Physical Model}\label{sec2}
	
\begin{figure}[htbp]
	\centering
	\includegraphics[bb=102 68 451 281, width=8.5cm, clip]{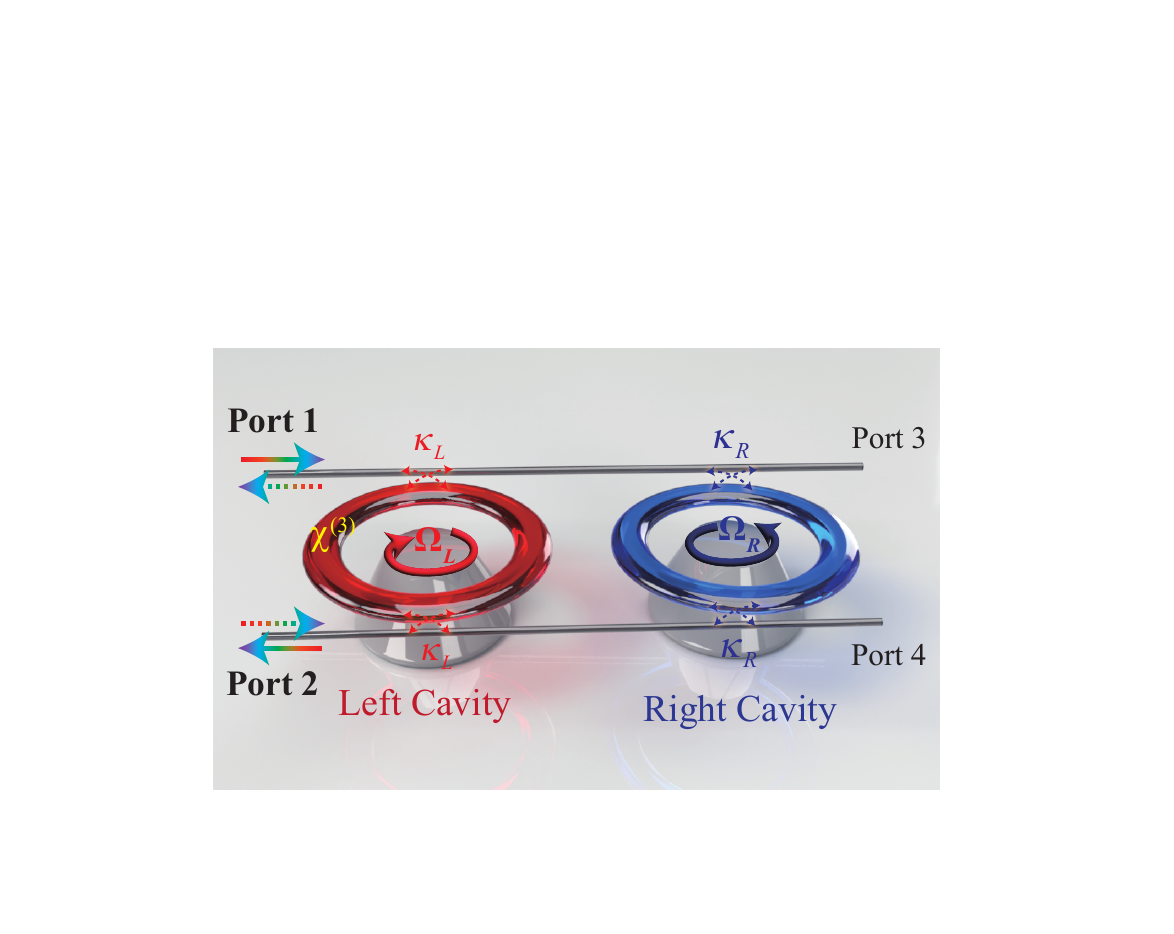}
	\caption{The schematic diagram of the optical system: A spinning nonlinear optical cavity (left) and a spinning linear optical cavity (right) are coupled to two common drop-filter waveguides (four ports) with coupling strengths (decay rates $\kappa_{L}$ and $\kappa_{R}$), and a weak driving field (frequency $\omega_d$) is input from Port $1$ or Port $2$.}
	\label{fig1}
\end{figure}
	
The optical system we consider is illustrated in Fig.~\ref{fig1}. 
A spinning nonlinear optical cavity (left) and a spinning linear optical cavity (right) are coupled to two common drop-filter waveguides (four ports) with coupling strengths (decay rates $\kappa_{L}$ and $\kappa_{R}$), and a weak driving field (frequency $\omega_d$) is input from Port $1$ or Port $2$.
In a frame rotating at the driving frequency $\omega_d$, the system can be described by the Hamiltonian ($\hbar = 1$)
\begin{eqnarray}\label{Eq1}
		H_{\sigma} &=& \Delta _{L,\sigma}a_{L,\sigma}^{\dag }a_{L,\sigma}+U a_{L,\sigma}^{\dag }a_{L,\sigma}^{\dag }a_{L,\sigma}a_{L,\sigma}\nonumber \\
		&&+\Delta _{R,\sigma}a_{R,\sigma}^{\dag }a_{R,\sigma}+J_{\mathrm{eff}}(a_{L,\sigma}^{\dag }a_{R,\sigma}+a_{R,\sigma}^{\dag }a_{L,\sigma}) \nonumber \\
		&&+i(\varepsilon _{L} a_{L,\sigma}^{\dag } -\varepsilon _{R} e^{i\theta }a_{R,\sigma}^{\dag }- {\rm H.c.}), 
\end{eqnarray}
where $\sigma={\rm cw}$ corresponds to the case that the weak driving field is input from Port $1$, and the clockwise modes ($a_{L,{\rm cw}}$ and $a_{R,{\rm cw}}$) are excited; $\sigma={\rm ccw}$ corresponds to the case that the weak driving field is input from Port $2$, and the courter-clockwise modes ($a_{L,{\rm ccw}}$ and $a_{R,{\rm ccw}}$) are excited. 
$\omega _{L}$ ($\omega _{R}$) is the resonance frequency of the left (right) cavity without spinning. 
The detunings are given by $\Delta _{L,{\rm cw}}=\Delta _{L}-\Delta_{F,L}$, $\Delta _{L,{\rm ccw}}=\Delta _{L}+\Delta_{F,L}$,  $\Delta _{R,{\rm cw}}=\Delta _{R}+\Delta_{F,R}$, $\Delta _{R,{\rm ccw}}=\Delta _{R}-\Delta_{F,R}$, $\Delta _{L}=\omega _{L}-\omega _{d} $ and $\Delta _{R}=\omega _{R}-\omega _{d}$. 	Here, we consider the case that the left (right) cavity rotates clockwise (counterclockwise) with an angular velocity $\Omega_L$ ($\Omega_R$), so the left (right) cavity experiences a Fizeau shift $\Delta _{F,L}$ ($\Delta _{F,R}$), with~\cite{Grigorii_2000}
\begin{equation}
		\Delta _{F,j}=\frac{n_{j}r_{j}\Omega _{j}\omega _{j}}{c}\left( 1-\frac{1}{%
			n_{j}^{2}}-\frac{\lambda }{n_{j}}\frac{dn_{j}}{d\lambda }\right),
\end{equation}%
where $n_{j}$ ($j=L,\,R$) is the refractive index, $r_{j}$ is the resonator radius of the cavity, $\lambda$ is the wavelength of the light in vacuum, $c$ is the speed of light in vacuum. 
The dispersion term $dn/d\lambda$ is originated from the relativistic effect and it is remarkably small (to $\thicksim 1\%$)~\cite{Maayani2018Natur,Grigorii_2000}.
$U=\hbar \omega _{L}^{2}cn_{2}/(n_{L}^{2}V_{\mathrm{eff}})$ is the Kerr interaction strength~\cite{Marin2017Natur}, where $n_{2}$ is the nonlinear refraction index, and $V_{\mathrm{eff}}$ is the effective mode volume. 
$\varepsilon _{L}= \sqrt{\kappa _{L}P_{\mathrm{in}}/(\hbar \omega _{d})}$ ($\varepsilon _{R}= \sqrt{\kappa _{R}P_{\mathrm{in}}/(\hbar \omega _{d})}$) is the driving amplitude of the left (right) cavity, and $P_{\mathrm{in}}$ is the driving power. $J_{\mathrm{eff}}=i\sqrt{\kappa _{L}\kappa _{R}}e^{i\theta }$ describes the waveguide-induced indirect coupling strength between the two cavities~\cite{PengZH2020PRA}, and $\theta$ is accumulated phase of light propagating from one cavity to another in the waveguides. Here, we consider $\theta=\pi/2$ so that the Hamiltonian (\ref{Eq1}) is a Hermitian.
	

In order to describe the transmission behaviors and quantum statistical properties for the photons transport from Port $i$ to Port $j$,
we define the optical transmission rate $T_{ji}$ and the equal-time second-order correlation function $g_{ji}^{(2)}(0)$ as
\begin{eqnarray}\label{4}
		T_{ji} &=& \frac{\left\langle a_{j,\mathrm{out}}^{\dag }a_{j,\mathrm{out}}\right\rangle }{\left\langle a_{i,\mathrm{in}}^{\dag }a_{i,\mathrm{in}%
			}\right\rangle },
\end{eqnarray}
and
\begin{equation}\label{5}
		g_{ji}^{(2)}(0)=\frac{\left\langle a_{j,\mathrm{out}}^{\dag }a_{j,\mathrm{out}}^{\dag }a_{j,\mathrm{out}}a_{j,\mathrm{out}}\right\rangle }{%
			\left\langle a_{j,\mathrm{out}}^{\dag }a_{j,\mathrm{out}}\right\rangle ^{2}},
\end{equation}
where $a_{i,\mathrm{in}}$ is the optical field input from Port $i$ and $a_{j,\mathrm{out}}$ is the optical field output from Port $j$.
In this paper, we focus on the nonreciprocal transmission between Port $1$ and $2$, with a weak driving field input from Port $1$ or Port $2$.
According to the input-output relations~\cite{gardiner1985}, the output fields from Port $1$ and $2$ ($a_{\mathrm{1,out}}$ and $a_{\mathrm{2,out}}$) can be expressed as
\begin{eqnarray}\label{8}
		a_{\mathrm{1,out}} &=&\sqrt{\kappa _{L}}a_{L\mathrm{,ccw}}-i\sqrt{%
			\kappa _{R}}a_{R\mathrm{,ccw}}+ia_{\mathrm{3,in}}, \\
		a_{\mathrm{2,out}} &=&\sqrt{\kappa _{L}}a_{L\mathrm{,cw}}-i\sqrt{%
			\kappa _{R}}a_{R\mathrm{,cw}}+ia_{\mathrm{4,in}}, \label{9}
\end{eqnarray}
where $a_{\mathrm{3,in}}$ ($a_{\mathrm{4,in}}$) is the input field from Port $3$ (Port $4$). Here, both $a_{\mathrm{3,in}}$ and $a_{\mathrm{4,in}}$ are vacuum fields.

Specifically, if a coherent field is input from Port $1$, i.e., $\left\langle a_{\mathrm{1,in}}\right\rangle=\sqrt{P_{\mathrm{in}}/(\hbar \omega _{d})}$ and $\left\langle a_{\mathrm{2,in}}\right\rangle=0$, then the transmission rate from Port $1$ to $2$ is given by
\begin{equation}\label{T21}
		T_{21}=T_{21}^{L}+T_{21}^{R}+T_{21}^{I},
\end{equation}
where
\begin{equation}
		T_{21}^{L}=\frac{\kappa _{L}\left\langle a_{L\mathrm{,cw}}^{\dag }a_{L\mathrm{,cw}}\right\rangle}{\left\langle a_{\mathrm{1,in}}^{\dag }a_{\mathrm{1,in}}\right\rangle }
\end{equation}  
corresponds to the photons passing through the left cavity,
\begin{equation}
		T_{21}^{R}=\frac{\kappa _{R}\left\langle a_{R\mathrm{,cw}}^{\dag}a_{R\mathrm{,cw}}\right\rangle}{\left\langle a_{\mathrm{1,in}}^{\dag }a_{\mathrm{1,in}}\right\rangle }
\end{equation}
corresponds to the photons passing through the right cavity,
and 
\begin{equation}
		T_{21}^{I}=\frac{-2\sqrt{\kappa _{L}\kappa _{R}}\mathrm{Re}\left[i\left\langle a_{L\mathrm{,cw}}^{\dag }a_{R\mathrm{,cw}}\right\rangle \right]}{\left\langle a_{\mathrm{1,in}}^{\dag }a_{\mathrm{1,in}}\right\rangle }
\end{equation}
is the interference term between the two paths.
In the meantime, the equal-time second-order correlation function $g_{21}^{(2)}(0)$ is given by
\begin{eqnarray}
		g_{21}^{(2)}(0) &=&\frac{1}{\left\langle a_{\mathrm{2,out}}^{\dag }a_{\mathrm{2,out}} \right\rangle ^{2}%
		}\,\Big(\kappa _{L}^{2}\left\langle a_{L\mathrm{,cw}}^{\dag }a_{L\mathrm{,cw}%
		}^{\dag }a_{L\mathrm{,cw}}a_{L\mathrm{,cw}}\right\rangle \nonumber \\
		&&+\kappa_{R}^{2}\left\langle a_{R\mathrm{,cw}}^{\dag }a_{R\mathrm{,cw}}^{\dag }a_{%
			R\mathrm{,cw}}a_{R\mathrm{,cw}}\right\rangle   \nonumber \\
		&&+4\kappa _{L}\kappa _{R}\left\langle a_{L\mathrm{,cw}}^{\dag }a_{R\mathrm{%
				,cw}}^{\dag }a_{L\mathrm{,cw}}a_{R\mathrm{,cw}}\right\rangle   \nonumber \\	
		&&-4\kappa _{L}\sqrt{\kappa _{L}\kappa _{R}}\,\mathrm{Re}\left[i\left\langle a_{L\mathrm{,cw}}^{\dag }a_{L\mathrm{,cw}}^{\dag }a_{L\mathrm{%
				,cw}}a_{R\mathrm{,cw}}\right\rangle \right]   \nonumber \\
		&&-4\kappa _{R}\sqrt{\kappa _{L}\kappa _{R}}\,\mathrm{Re}\left[i\left\langle a_{L\mathrm{,cw}}^{\dag }a_{R\mathrm{,cw}}^{\dag }a_{R\mathrm{%
				,cw}}a_{R\mathrm{,cw}}\right\rangle \right]  \nonumber \\
		&&-2\kappa _{L}\kappa _{R}\,\mathrm{Re}\left[\left\langle a_{%
			L\mathrm{,cw}}^{\dag }a_{L\mathrm{,cw}}^{\dag }a_{R\mathrm{,cw}}a_{R\mathrm{%
				,cw}}\right\rangle \right]\Big),
\end{eqnarray}
which indicates that $g_{21}^{(2)}(0)$ not only relates to the self-correlation of the photons in the two cavities (the first two terms), but also depends on the cross-correlation of the photons between the two cavities (the last four terms). 
	
Similarly, if a coherent field is input from Port $2$, i.e. $\left\langle a_{\mathrm{2,in}}\right\rangle=\sqrt{P_{\mathrm{in}}/(\hbar \omega _{d})}$ and $\left\langle a_{\mathrm{1,in}}\right\rangle=0$, then the transmission rate from Port $2$ to $1$ is given by
\begin{equation}\label{T12t}
		T_{12}=T_{12}^{L}+T_{12}^{R}+T_{12}^{I},
\end{equation}
with
\begin{equation}
		T_{12}^{L}=\frac{\kappa _{L}\left\langle a_{L\mathrm{,ccw}}^{\dag }a_{L\mathrm{,ccw}}\right\rangle}{\left\langle a_{\mathrm{2,in}}^{\dag }a_{\mathrm{2,in}}\right\rangle }
\end{equation}  
for photons passing through the left cavity,
\begin{equation}\label{T12}
		T_{12}^{R}=\frac{\kappa _{R}\left\langle a_{R\mathrm{,ccw}}^{\dag}a_{R\mathrm{,ccw}}\right\rangle}{\left\langle a_{\mathrm{2,in}}^{\dag }a_{\mathrm{2,in}}\right\rangle }
\end{equation}
for photons passing through the right cavity,
and 
\begin{equation}
		T_{12}^{I}=\frac{-2\sqrt{\kappa _{L}\kappa _{R}}\mathrm{Re}\left[i\left\langle a_{L\mathrm{,ccw}}^{\dag }a_{R\mathrm{,ccw}}\right\rangle \right]}{\left\langle a_{\mathrm{2,in}}^{\dag }a_{\mathrm{2,in}}\right\rangle }
\end{equation}
for the interference between the two paths.
In the meantime, we have the equal-time second-order correlation function
\begin{eqnarray}
		g_{12}^{(2)}(0) &=&\frac{1}{\left\langle a_{\mathrm{1,out}}^{\dag }a_{\mathrm{1,out}} \right\rangle ^{2}%
		} \Big(\kappa _{L}^{2}\left\langle a_{L\mathrm{,ccw}}^{\dag }a_{L\mathrm{,ccw}%
		}^{\dag }a_{L\mathrm{,ccw}}a_{L\mathrm{,ccw}}\right\rangle \nonumber \\
		&&+\kappa_{R}^{2}\left\langle a_{R\mathrm{,ccw}}^{\dag }a_{R\mathrm{,ccw}}^{\dag }a_{%
			R\mathrm{,ccw}}a_{R\mathrm{,ccw}}\right\rangle   \nonumber \\
		&&+4\kappa _{L}\kappa _{R}\left\langle a_{L\mathrm{,ccw}}^{\dag }a_{R\mathrm{%
				,ccw}}^{\dag }a_{L\mathrm{,ccw}}a_{R\mathrm{,ccw}}\right\rangle   \nonumber
		\\
		&&-4\kappa _{L}\sqrt{\kappa _{L}\kappa _{R}}\,\mathrm{Re}\left[i\left\langle a_{L\mathrm{,ccw}}^{\dag }a_{L\mathrm{,ccw}}^{\dag }a_{L\mathrm{%
				,ccw}}a_{R\mathrm{,ccw}}\right\rangle \right]   \nonumber \\
		&&-4\kappa _{R}\sqrt{\kappa _{L}\kappa _{R}}\,\mathrm{Re}\left[i\left\langle a_{L\mathrm{,ccw}}^{\dag }a_{R\mathrm{,ccw}}^{\dag }a_{R\mathrm{%
				,ccw}}a_{R\mathrm{,ccw}}\right\rangle \right] \nonumber \\
		&&-2\kappa _{L}\kappa _{R}\,\mathrm{Re}\left[\left\langle a_{%
			L\mathrm{,ccw}}^{\dag }a_{L\mathrm{,ccw}}^{\dag }a_{R\mathrm{,ccw}}a_{%
			R\mathrm{,ccw}}\right\rangle \right] \Big),
\end{eqnarray}
which is used to described the statistical properties of the photons transport from Port $2$ to $1$.

The quantum dynamics of the system is governed by the master equation~\cite{carmichael1993}
\begin{equation}
		\dot{\rho}_{\sigma} =-i[H_{\sigma},\rho_{\sigma}]+2\kappa _{1}L[a_{L,\sigma}]\rho_{\sigma} +2\kappa_{2}L[a_{R,\sigma}]\rho_{\sigma},
\end{equation}
where ${\rho}_{\sigma}$ ($\sigma={\rm cw,\, ccw}$) is the density operator and $L[o]\rho_{\sigma} =o\rho_{\sigma} o^{\dag }-(o^{\dag }o\rho_{\sigma} +\rho_{\sigma} o^{\dag }o)/2$ denotes a Lindbland term for an operator $o$. The transmission rates and second-order correlation functions can be obtained by solving the master equation numerically. 
	
In this paper, we choose the experimentally accessible parameters as~\cite{huetMilli2016,pavlovSolitonDual2017,WhisperingGalleryMode2011,schuster2008,shen2016,spillane2005,vahala2003,zielinska2017}: $\lambda_L=\lambda_R=1550$ nm, $Q_L=Q_R=2.5\times 10^{9}$, $V_{\mathrm{eff}}=147\,\,\mu \mathrm{m^{3}}$, $n_{2}=3\times 10^{-14}\,\,\mathrm{m^{2}}/\mathrm{W}$, $r_{L}=r_{R}=30\,\,\mu \mathrm{m}$, $n_{L}=n_{R}=1.4 $, $P_\mathrm{in}=0.2\,\,\mathrm{fW}$,  $\Omega_{L}/2\pi=\Omega_{R}/2\pi=9.4\,\,\mathrm{kHz}$. For simplicity, we set $\Delta_L=\Delta_R=\Delta$ and $\kappa_L=\kappa_R=\kappa$.
In the microring resonators, $Q$ is usually $10^9-10^{12}$~\cite{huetMilli2016,pavlovSolitonDual2017,WhisperingGalleryMode2011}, and $V_{\mathrm{eff}}$ is typically $10^2-10^4~\mu \mathrm{m^{3}}$~\cite{spillane2005,vahala2003}. 
The Kerr coefficient can be $n_{2}\thicksim  10^{-14}$ for materials with potassium titanyl phosphate~\cite{zielinska2017}.
In addition, the left (right) cavity rotating rotates clockwise (counterclockwise) with frequency $\Omega_{L}/2\pi=9.4\,\,\mathrm{kHz}$ ($\Omega_{R}/2\pi=9.4\,\,\mathrm{kHz}$), leading to the Fizeau shifts of $\Delta_{F,L\mathrm{}}=\Delta_{F,R\mathrm{}}=\Delta_{F}\approx 20\kappa$.
The resonator with a radius of $1.1$ mm can rotate at an rotation frequency of $6.6$ kHz~\cite{Maayani2018Natur}, and a higher angular frequency can be achieved for a smaller object, such as the single $100$ nm particles with a rotation frequency of GHz has been observed experimentally~\cite{ahn2018,Jin2021PRJ}.
	
\begin{figure*}[htbp]
		\centering
		\includegraphics [bb=430 630 1419 1145, width=17 cm, clip]{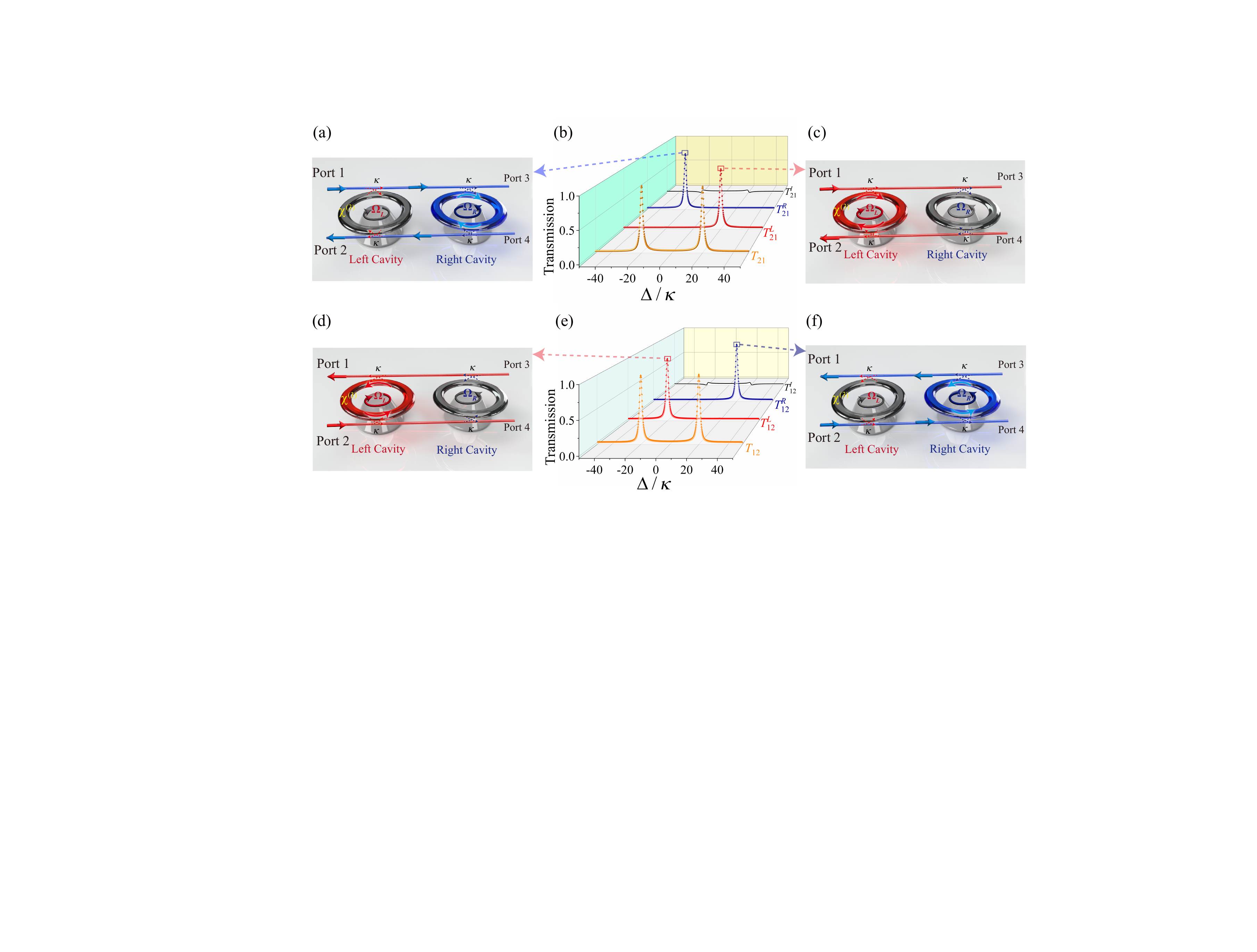}                         
		\caption{Spatially separated transmission scheme. 
			(b) The transmission rates $T_{21}$, $T_{21}^{L}$,  $T_{21}^{R}$, and  $T_{21}^{I}$ versus the detuning $\Delta/\kappa$; 
			(e) the transmission rates $T_{12}$, $T_{12}^{L}$,  $T_{12}^{R}$, and  $T_{12}^{I}$ versus the detuning $\Delta/\kappa$.
			(a) and (c) The schematic diagram of the paths for photons transport from Port 1 to 2;
			(d) and (f) The schematic diagram of the paths for photons transport from Port 2 to 1.
		}
		\label{fig2}
\end{figure*}

\section{spatially separated transmission}\label{sec3}

In this section, we introduce a spatially separated transmission scheme that the photons transport from Port 1 to 2 and the ones transport from Port 2 to 1 take different paths (cavities).
Under the conditions that the two spatially separated cavities work on the same resonance frequency ($\omega_L=\omega_R$) and they spin in opposite directions with the same rotating frequency ($\Omega_L=\Omega_R$), we find that the photon transmission between Port 1 and 2 is reciprocal, i.e., the transmission rate for photons transport from Port 1 to 2 is the same as the one for photons transport from Port 2 to 1 ($T_{21}\approx T_{12}$) in the whole spectra, as shown in Figs.~\ref{fig2}(b) and \ref{fig2}(e).
However, the photons input from different ports take different paths (cavities), which is the ingredient for purely quantum nonreciprocity (no classical nonreciprocity) we will discuss in the next section.
	
According to Eq.~(\ref{T21}), the transmission spectrum for the photons transport from Port 1 to 2 can be divided into three parts: $T_{21}^{L}$, $T_{21}^{R}$, and $T_{21}^{I}$, as shown in Fig.~\ref{fig2}(b).
There are two resonance peaks in the transmission spectra $T_{21}$, i.e., $\Delta\approx \pm 20\kappa$. The resonance transmission around the detuning $\Delta=20\kappa$ corresponds to the case that the photons transport from Port 1 to 2 by passing through the left cavity with $T_{21}^{L}\approx 1$ [Fig.~\ref{fig2}(c)]. In contrast, resonance transmission around the detuning $\Delta=-20\kappa$ corresponds to the case that photons transport from Port 1 to 2 by passing through the right cavity with $T_{21}^{R}\approx 1$  [Fig.~\ref{fig2}(a)].
Moreover, the interference term $T_{21}^{I}$ is very small and can be ignored, due to the large detuning between the resonance frequencies for the two optical cavities spinning in the opposite direction, $|\Delta _{L,{\rm cw}}-\Delta _{R,{\rm cw}}|\approx 40 \kappa $.
	
The transmission spectrum for the photons transport from Port 2 to 1 can also be divided into three parts [Eq.~(\ref{T12t})]: $T_{12}^{L}$, $T_{12}^{R}$, and $T_{12}^{I}$, as shown in Fig.~\ref{fig2}(e).
There are also two resonance peaks in the transmission spectra $T_{12}$, i.e., $\Delta\approx \pm 20\kappa$. Different from the case of the photons transport from Port 1 to 2, the resonance transmission around the detuning $\Delta=20\kappa$ corresponds to the case that the photons transport from Port 2 to 1 by passing through the right cavity with $T_{12}^{R}\approx 1$ [Fig.~\ref{fig2}(f)]. In contrast, resonance transmission around the detuning $\Delta=-20\kappa$ corresponds to the case that photons transport from Port 2 to 1 by passing through the left cavity with $T_{12}^{L}\approx 1$  [Fig.~\ref{fig2}(d)].
The interference term $T_{12}^{I}$ is also very small and can be ignored due to the large detuning between the resonance frequencies for the two optical cavities spinning in the opposite direction.
	
Based on the above discussion, we have confirmed that the photon transmission between Port 1 and 2 is reciprocal, i.e., $T_{21}\approx T_{12}$, but the photons pass through different paths when they transport in different directions.
To be more specific, at detuning $\Delta\approx 20\kappa$, the photons transport from Port 1 to 2 by passing through the left cavity with $T_{21}^{L}\approx 1$, while the photons transport from Port 2 to 1 by passing through the right cavity with $T_{12}^{R}\approx 1$; in contrast, at detuning $\Delta\approx -20\kappa$, the photons transport from Port 1 to 2 by passing through the right cavity with $T_{21}^{R}\approx 1$, while the photons transport from Port 2 to 1 by passing through the left cavity with $T_{12}^{L}\approx 1$. Such a spatially separated transmission scheme provides us an ideal platform to realize purely quantum nonreciprocity (no classical nonreciprocity) by considering the nonlinear interaction in one of the paths (cavities).

\section{Purely Quantum Nonreciprocity}\label{sec4}

\begin{figure}[htbp]
		\includegraphics[bb=160 470 353 642, width=8.5cm, clip]{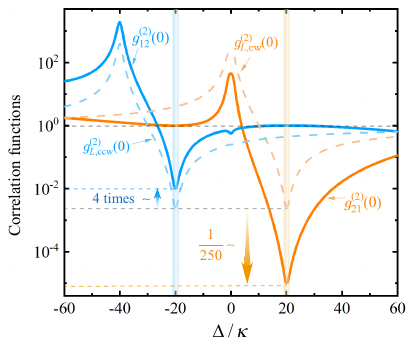}
		\centering
		\caption{
			The second-order correlation functions $g_{21}^{(2)}(0)$, $g_{12}^{(2)}(0)$, $g_{L,{\rm cw}}^{(2)}(0)$, and $g_{L,{\rm ccw}}^{(2)}(0)$ versus the detuning $\Delta/\kappa$.
		}
		\label{fig3}
\end{figure}

Based on the spatially separated transmission scheme discussed above, here we show that purely quantum nonreciprocity (nonreciprocal photon blockade) can be observed by considering the Kerr nonlinear interaction in one of the cavities (left cavity).
We can predict that if a coherent field (weak driving field) is input from one of the ports, then we will observe photon blockade in the output field if the photons pass through the left cavity due to the strong Kerr nonlinear interaction,
or we still obtain a coherent field in the output field if the photons pass through the right (linear) cavity.

The equal-time second-order correlation functions $g_{21}^{(2)}(0)$ and $g_{12}^{(2)}(0)$ are shown in Fig.~\ref{fig3}.
We achieve the nonreciprocal photon blockade in the parameter regions around the detuning $\Delta=\pm 20\kappa$ for high transmission rate $T_{21}=T_{12}\approx1$.
According to Figs.~\ref{fig2}(a)-\ref{fig2}(c), the photons transport from Port 1 to 2 passing through the left (nonlinear) cavity at $\Delta= 20\kappa$, thus we obtain sub-Poissonian statistical distribution ($g^{(2)}_{21}(0)\ll 1$) in the output field. In contrast, the photons transport from Port 2 to 1 passing through the right (linear) cavity at $\Delta= 20\kappa$, thus we obtain a Poisson statistical distribution ($g^{(2)}_{12}(0) \approx 1$) in the output field.
On the contrary, at $\Delta=- 20\kappa$, the photons transport from Port 1 to 2 passing through the right cavity and we obtain a coherent output field ($g^{(2)}_{21}(0)\approx  1$); the photons transport from Port 2 to 1 passing through the left cavity and we obtain single photons ($g^{(2)}_{12}(0) \ll1$) in the output field.
	
Surprisingly, the nonreciprocal photon blockade is enhanced nonreciprocally, i.e., the minimal value of $g^{(2)}_{21}(0)$ is much smaller than that of  $g^{(2)}_{12}(0)$.
For comparison, we also introduce the equal-time second-order correlation function of the photons in the left cavity as
\begin{equation}
		g_{L,\sigma}^{(2)}(0)=\frac{\left\langle a_{L,\sigma}^{\dag }a_{L,\sigma}^{\dag }a_{L,\sigma}a_{L,\sigma}\right\rangle}{\left\langle a_{L,\sigma}^{\dag }a_{L,\sigma}\right\rangle^{2}},
\end{equation}
where $\sigma={\rm cw}$ or ${\rm ccw}$. To be clear, $g_{L,\sigma}^{(2)}(0)$ are also shown (dashed curves) in Fig.~\ref{fig3}.
The minimal value of $g_{12}^{(2)}(0)$ is about $4$ time that of $g_{L,{\rm ccw}}^{(2)}(0)$ and
the minimal value of $g_{21}^{(2)}(0)$ is about $1/250$ that of $g_{L,{\rm cw}}^{(2)}(0)$. That means, in comparison with the photons in the (left) cavity, the photon blockade in the output fields is suppressed when the photons transport from Port 2 to 1, but significantly enhanced when the photons transport from Port 1 to 2.
As the minimal values of $g_{L,{\rm cw}}^{(2)}(0)$ and $g_{L,{\rm ccw}}^{(2)}(0)$ are almost the same,
thus we find that the minimal value of $g^{(2)}_{21}(0)$ is about three orders of magnitude smaller than that of  $g^{(2)}_{12}(0)$.
The mechanism for the photon blockade nonreciprocal enhancement is discussed in the next section.

\section{Mechanism for Photon Blockade nonreciprocal enhancement}\label{sec5}
	
In order to understand the origin of photon blockade nonreciprocal enhancement predicted above, we derive the analytical expressions of
the equal-time second-order correlation functions based on the Schr\"{o}dinger equation.
	
The wave function of the system can be expanded on the Fock-state basis $\left\vert n_{L}n_{R}\right\rangle $, where $n_{L}$ and $n_{R}$ denote the number of photons in the left and right cavity, respectively. In the limit of weak driving field, the system can be truncated up to at most two photons, i.e, $n_{L}+n_{R}\leq 2$. In the truncated space, the state of the system can be described in the following form,
\begin{eqnarray}\label{tai}
		\left\vert \varphi (t)\right\rangle  &=&C_{00}\left\vert
		00\right\rangle +C_{10}\left\vert 10\right\rangle
		+C_{01}\left\vert 01\right\rangle   \nonumber \\
		&&+C_{11}\left\vert 11\right\rangle +C_{20}\left\vert
		20\right\rangle +C_{02}\left\vert 02\right\rangle ,
\end{eqnarray}
where $C_{n_Ln_R}$ is the probability amplitude of the Fock state $\left\vert n_{L}n_{R}\right\rangle $.
According to the quantum-trajectory method~\cite{RevModPhys.70.101}, the system (with decay rate $2\kappa$ for each cavity) is governed by a non-Hermitian Hamiltonian
\begin{equation}\label{Heff}
		\tilde{H}_{\sigma }=H_{\sigma}-i \kappa a_{L,\sigma}^{\dag }a_{L,\sigma}-i \kappa a_{R,\sigma}^{\dag }a_{R,\sigma},
\end{equation}
where $\sigma={\rm cw}$ or ${\rm ccw}$.
Substituting the wave function (\ref{tai}) and non-Hermitian Hamiltonian (\ref{Heff}) into the Schr\"{o}dinger equation $i\partial \left\vert \varphi (t)\right\rangle /\partial t=\tilde{H}_{\sigma } \left\vert \varphi (t)\right\rangle$, we can obtain the analytical expressions of the coefficients $C_{n_{L}n_{R}}$ in the steady state.
	
Based on the analytical expressions of the coefficients $C_{n_{L}n_{R}}$ in the steady state, the second-order correlation function of the field output from Port $j$ (input from Port $i$) can be written as
\begin{eqnarray}\label{g2out}
		g_{ji}^{(2)}(0) &\approx &\frac{2}{\left\vert C_{10}\right\vert
			^{4}}\Big\{\left\vert C_{20}-\sqrt{2}iC_{11}\right\vert ^{2}+\left\vert
		C_{02}\right\vert ^{2}  \nonumber \\
		&&-2\mathrm{Re}\left[ \left( C_{20}-\sqrt{2}iC_{11}\right)
		C_{02}^{\ast }\right] \Big\},
\end{eqnarray}
where $ji=21$ or $12$.
Correspondingly, the second-order correlation function of the photons in the left cavity is given by
\begin{eqnarray}\label{g2l}
		g_{L,\sigma}^{(2)}(0) &\approx &\frac{2 \left\vert C_{20}\right\vert ^{2}}{\left\vert C_{10}\right\vert^{4}}
\end{eqnarray}
with $\sigma={\rm cw}$ or ${\rm ccw}$.

\begin{figure}[tbp]
	\centering
	\includegraphics[bb=32 900 1327 1572, width=8.5 cm, clip]{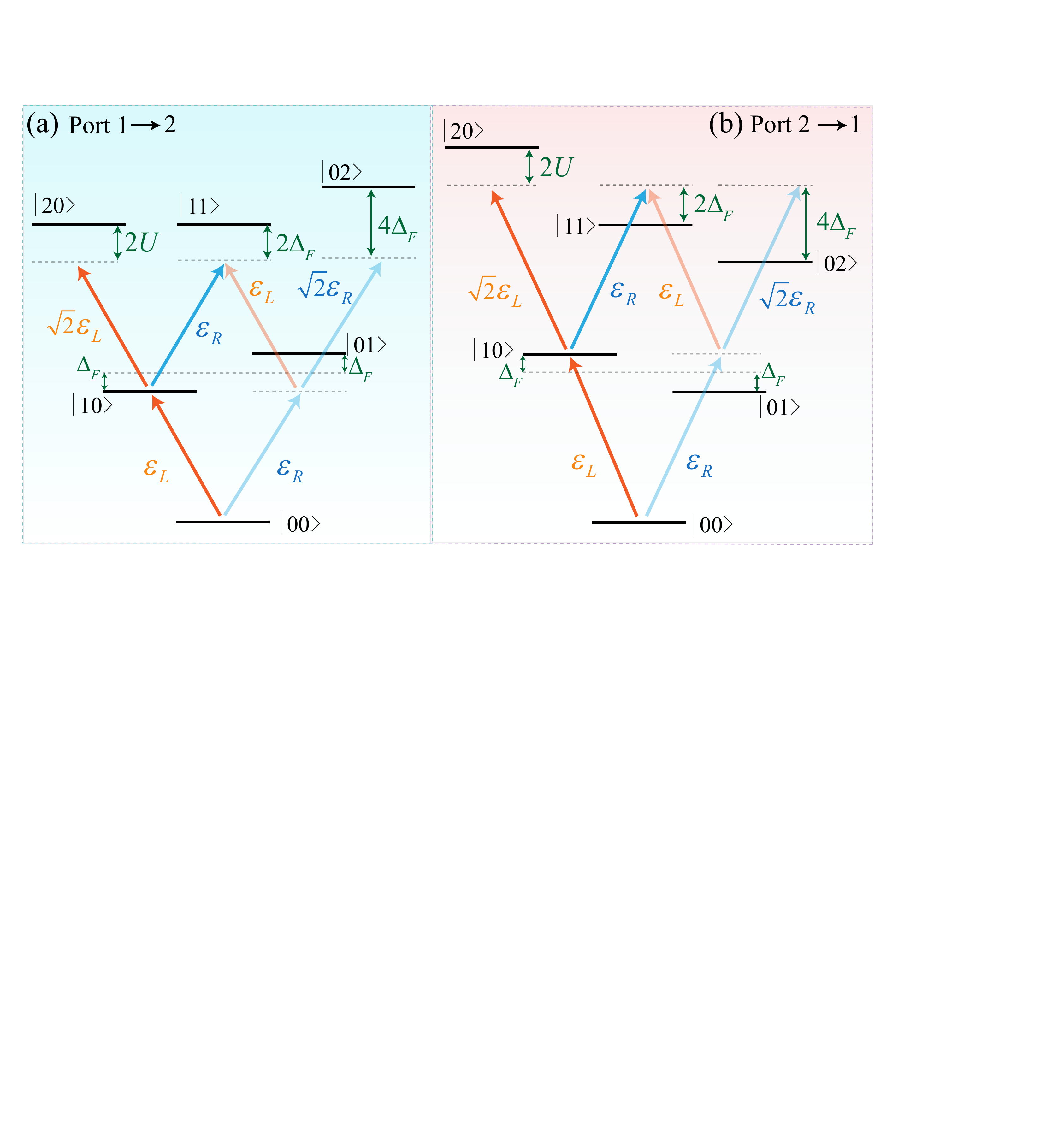}
	\caption{Schematic energy spectrum of the system in the low-excitation
			subspace. (a) The photons transport from Port 1 to 2; (b) the photons transport from Port 2 to 1.}
	\label{fig4}
\end{figure}

To find the optimal conditions for photon blockade nonreciprocal enhancement, we show the energy levels of the system in Fig.~\ref{fig4}.
According to the previous work~\cite{Imamoglu1997PRL}, photon blockade should be observed under the conditions that the single-photon state is driven resonantly, i.e., resonant driving between levels $|00\rangle$ and $|10\rangle$, and the two-photon states are driven non-resonantly.
Thus, we obtain one of the optimal conditions for photon blockade: $\Delta_{L,{\rm cw}}=0$ ($\Delta=\Delta_F$) for the photons transport from Port $1$ to $2$ [Fig.~\ref{fig4}(a)]; $\Delta_{L,{\rm ccw}}=0$ ($\Delta=-\Delta_F$) for the photons transport from Port $2$ to $1$ [Fig.~\ref{fig4}(b)].
Moreover, by comparing Eqs.~(\ref{g2out}) and (\ref{g2l}), photon blockade enhancement should be observed under the conditions $\left\vert C_{2,0}\right\vert \approx \sqrt{2}\left\vert C_{1,1}\right\vert \gg\left\vert C_{0,2}\right\vert$, so that the term $ (C_{20}-\sqrt{2}iC_{11})$ can be canceled out by destructive interference.
The condition $\left\vert C_{2,0}\right\vert \approx \sqrt{2}\left\vert C_{1,1}\right\vert$
is achieved when the transition frequencies of $|10\rangle\rightarrow |20\rangle$ and $|10\rangle\rightarrow |11\rangle$ are the same, i.e., $U=\Delta_F$.
Based on the above analysis, we find that the optimal conditions for photon blockade nonreciprocal enhancement are $\Delta=\Delta_F=U$, which are consistent well with the numerical results shown in Fig.~\ref{fig3}.

\begin{figure}[tbp]
	\includegraphics[bb=188 360 494 590, width=8cm, clip]{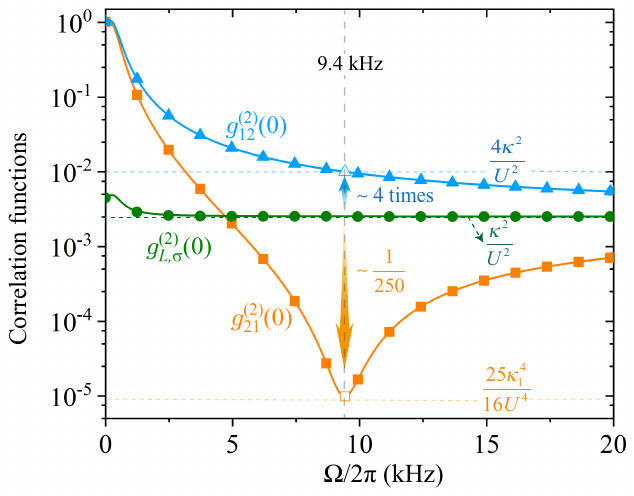}
	\centering
	\caption{
			The second-order correlation functions $g_{21}^{(2)}(0)$, $g_{12}^{(2)}(0)$, and $g_{L,\sigma}^{(2)}(0)$ versus the rotation frequency $\Omega/2\pi$. Here, $g_{21}^{(2)}(0)$ is plotted with the detuning $\Delta=\Delta_{F}$,  $g_{21}^{(2)}(0)$ is plotted with the detuning $\Delta=-\Delta_{F}$.}
	\label{fig5}
\end{figure}

To verify the optimal conditions ($\Delta=\Delta_F=U$) for photon blockade nonreciprocal enhancement, we show $g_{21}^{(2)}(0)$ and $g_{L,{\rm cw}}^{(2)}(0)$ ($g_{12}^{(2)}(0)$ and $g_{L,{\rm ccw}}^{(2)}(0)$), as functions of the rotation frequency $\Omega/2\pi$, under the resonant condition $\Delta=\Delta_F$ ($\Delta=-\Delta_F$), in Fig.~\ref{fig5}. 
When the two cavities are not spinning or spinning with a low angular velocity $\Omega/2\pi< 5$ kHz, we have $g_{12}^{(2)}(0)>g_{21}^{(2)}(0)>g_{L,{\rm cw}}^{(2)}(0)=g_{L,{\rm ccw}}^{(2)}(0)$.
But with the increase of the angular velocity $\Omega$, $g_{21}^{(2)}(0)$ decreases rapidly and the system comes into the regime of photon blockade nonreciprocal enhancement, i.e., $g_{12}^{(2)}(0)>g_{L,{\rm cw}}^{(2)}(0)=g_{L,{\rm ccw}}^{(2)}(0)>g_{21}^{(2)}(0)$.
Most notably, $g_{21}^{(2)}(0)$ reaches its minimal value at the angular velocity $\Omega/2\pi \approx 9.4$ kHz, i.e., $\Delta_F\approx U$, and it is about three orders of magnitude smaller than that of  $g^{(2)}_{12}(0)$.
This indicates a purely nonreciprocal photon blockade with direction-dependent enhancement in the spinning cavities, and such effect has not been revealed previously.
	
Now, we focus on the optimal conditions for photon blockade nonreciprocal enhancement, i.e., $\Delta_{L,{\rm cw}}=0$ ($\Delta = \Delta _{F}= U$) for the photons transport from Port $1$ to $2$ and $\Delta_{L,{\rm ccw}}=0$ ($\Delta = -\Delta _{F}= -U$) for the photons transport from Port $2$ to $1$.
Under these conditions, the probability amplitudes for two-photon states in the steady state are approximately given by
\begin{equation}\label{24}
		C_{20} \approx \frac{-\varepsilon _{L}^{2}\left[ \left( 2U^{2}-\kappa^{2}\right) i\pm 4U\kappa\right] }{2\sqrt{2}U^{3}\kappa},
\end{equation}
\begin{equation} \label{25}
		C_{11} \approx \frac{\varepsilon _{L}^{2}\left[ iU\kappa+\left(
			\kappa ^{2}\mp 2U^{2}\right) \right] }{4U^{3}\kappa},
\end{equation}
\begin{equation}\label{26}
		C_{02} \approx \frac{-\varepsilon _{L}^{2}}{4\sqrt{2}U^{2}} ,
\end{equation}
where, the plus sign ($+$) in Eq.~(\ref{24}) and the minus sign ($-$) in Eq.~(\ref{25}) denote the case that the photons transport from Port 1 to 2, and the minus sign ($-$) in Eq.~(\ref{24}) and the plus sign ($+$) in Eq.~(\ref{25}) denote the case that the photons transport from Port 2 to 1.
In the strong nonlinear regime ($U\gg \kappa$), we have $C_{20}\approx \sqrt{2}iC_{11} $ for photons transport from Port 1 to 2, and they are canceled out by destructive interference, with $C_{20}-\sqrt{2}iC_{11} \approx 0$. 
Meanwhile, we have $C_{20}\approx- \sqrt{2}iC_{11} $ for photons transport from Port 2 to 1, and there is constructive interference between them, with $C_{20}-\sqrt{2}iC_{11} \approx 2C_{20}$. 
Based on the approximate expressions of the probability amplitudes, the second-order correlation functions are obtained
analytically as
\begin{eqnarray}
		g_{21}^{(2)}(0) &\approx& \frac{25\kappa^{4}}{16U^{4}}, \\ \label{27}
		g_{12}^{(2)}(0) &\approx&  4\frac{\kappa^{2}}{U^{2}}, \\ \label{28}
		g_{L,\sigma}^{(2)}(0) &\approx& \frac{\kappa^{2}}{U^{2}}. \label{29}
\end{eqnarray}
The analytical expressions of $g_{12}^{(2)}(0)$, $g_{21}^{(2)}(0)$, and $g_{L,\sigma}^{(2)}(0)$ [Eqs.~(\ref{27})-(\ref{29})] are shown (dashed lines) in Fig.~\ref{fig5}.
It is clear that we have $g_{12}^{(2)}(0)\approx 4g_{L,\sigma}^{(2)}(0)$ and $g_{21}^{(2)}(0)\approx g_{L,\sigma}^{(2)}(0)/256$, which are consistent well with the numerical results in Figs.~\ref{fig3} and \ref{fig5}.
It is worth mentioning that we have $g_{21}^{(2)}(0)\propto (\kappa/U)^4$ and $g_{L,\sigma}^{(2)}(0)\propto (\kappa/U)^2$, i.e., the spinning cavities can be used to achieve scaling enhancement of photon blockade~\cite{liu2024scaling}.
	
\section{conclusion}\label{sec6}
In conclusion, we have proposed a spatially separated transmission scheme that the photons transport in different directions take different paths, in an optical system consisting of two spinning cavities coupled indirectly by two common drop-filter waveguides.
Based on the spatially separated transmission scheme, we demonstrated a purely quantum nonreciprocity (nonreciprocal photon blockade) by considering the Kerr nonlinear interaction in one of the cavities (left cavity).
Even more interestingly, the nonreciprocal photon blockade is enhanced nonreciprocally, i.e., the nonreciprocal photon blockade is enhanced when the photons transport in one direction but suppressed in the reverse direction.
We have identified that the nonreciprocal enhancement of the photon blockade is induced by the destructive or constructive interference between two paths for two photons passing through the whole system. 
The spinning cavities may also work when they contain other nonlinear interactions, such as second-order nonlinearity~\cite{WangK2019PRA,ShenHZ2020PRA,GouC2023PRA}, optomechanical interaction~\cite{LiBJ2019PRJ,LiuYM2023OE,Shang_2021}, or atom-cavity interaction~\cite{XueW2020OL,JingYW2021PRA,Wang_2021EPL,LiuYM2023PRA,ZhangW2023SCPMA}, to achieve more quantum nonreciprocal effects, e.g., nonreciprocal entanglement~\cite{JiaoYF2020PRL,JiaoYF2022PRAPP,Ren2022OL,ChenJ2023PRB} and nonreciprocal squeezing~\cite{ChenSS2021ADP,Zhao2024OL,GuoQ2023PRA,wang2024quantum}.
Moreover, the spatially separated transmission scheme may also be extended to study other nonreciprocal effects, such as nonreciprocal cooling~\cite{Habraken_2012,XuH2019Natur,LaiDG2020PRA}, nonreciprocal lasing~\cite{JiangY2018PRAPP,Heras2021PRA,XuY2021PRA,XuY2021OL,LuTX2024SCPMA}, and nonreciprocal (topology) energy band by considering a array of cavities~\cite{Bahari2017Sci,Seif2018APS,HuN2023PRA,QinH2024PRA}.
	
\begin{acknowledgments}
		This work is supported by the National Natural Science Foundation of China (Grants No.~12064010 and  No.~12247105),
		the science and technology innovation Program of Hunan Province (Grant No.~2022RC1203), 
		and Hunan provincial major sci-tech program (Grant No.~2023ZJ1010).
\end{acknowledgments}

\bibliography{ref} 


\end{document}